\documentclass[%
 amsmath,amssymb,preprint,
prb,
floatfix,
]{revtex4-2}

\usepackage[version=3]{mhchem} 
\usepackage{tabu}
\usepackage{amsmath}
\usepackage{amssymb}
\usepackage{graphicx}
\usepackage{xcolor}
\usepackage{gensymb}
\usepackage{siunitx}
\usepackage{dblfloatfix} 

\begin{document}

\title{Indirect Mechanism of Au adatom Diffusion on the Si(100) Surface}

\author{Alejandro Pe\~{n}a-Torres}
\affiliation
{Science Institute of the University of Iceland, VR-III, 107 Reykjav\'{\i}k, Iceland}
\author{Abid Ali}
\affiliation
{Science Institute of the University of Iceland, VR-III, 107 Reykjav\'{\i}k, Iceland}
\author{Michail Stamatakis}
\affiliation
{Department of Chemical Engineering, University College London, Roberts Building, Torrington Place, London WC1E 7JE, UK }
\author{Hannes J\'onsson}
\affiliation
{Faculty of Physical Sciences, University of Iceland, VR-III, 107 Reykjav\'{\i}k, Iceland}
\affiliation{Department of Applied Physics, Aalto University, FI-00076 Espoo, Finland}







\begin{abstract}
Calculations of the diffusion of a Au adatom on the dimer reconstructed Si(100)-2$\times$1 surface reveal an interesting mechanism that differs significantly from a direct path between optimal binding sites, which are located in between dimer rows. 
Instead, the active diffusion mechanism involves promotion of the adatom to higher energy sites on top of a dimer row and then fast migration along the row, visiting {\it ca.} a hundred sites at room temperature, before falling back down into an optimal binding site. 
This top-of-row mechanism becomes more important the lower the temperature is. 
The calculations are carried out by finding minimum energy paths on the energy surface obtained from density functional theory within the PBEsol functional approximation followed by kinetic Monte Carlo simulations of the diffusion over a range of temperature from 200 K to 900 K. While the activation energy for the direct diffusion mechanism is calculated to be 0.84 eV, the effective activation energy for the indirect mechanism is on average 0.56 eV.
\end{abstract}

\maketitle

\section{Introduction}

The formation of metallic nanostructures on solid surfaces has become the focus of various types of research and technological applications~\cite{zhang2013plasmonic} as they can have interesting properties, such as optical, electronic and catalytic  \cite{link2000shape,roduner2006size}. Gold nanostructures are often of particular interest because
of their 
stability
and silicon surfaces represent a natural choice for a substrate because of its widespread use in electronic applications. 
For example, gold nanoparticles formed on a silicon surface can be used 
as metal catalysts in the synthesis of one-dimensional nanostructures such as carbon nanotubes~\cite{bhaviripudi2007cvd} and silicon nanowires~\cite{hannon2006influence}. 
They have also been found to display interesting optical properties~\cite{minoda2005anomalous} and to form mesoscopic structures \cite{choi2008band}. Moreover, the presence of a gold layer on silicon surface has proven to play an important role in the growth mechanism of silicon oxide \cite{leclerc1992nonlinear} and other materials \cite{cheng2008effects}.



An understanding of the interaction between Au atoms and the Si surface as well as the initial stages of Au nanostructure formation is, 
therefore, of considerable importance. 
Epitaxial growth of gold islands and overlayers on Si(100) have been studied by scanning and 
high-resolution transmission electron microscopy, electron diffraction 
and grazing-incidence X-ray diffraction \cite{kim1996growth,Piscopiello08,Wolz14}.
More detailed information about the Au/Si(100) interaction has been obtained from the low temperature
scanning tunneling microscopy (STM) studies of Chiaravalloti {\it et al.} where the binding sites of Au adatoms on the silicon surface could be identified \cite{chiaravalloti2009stm} both in between the Si dimer rows of the reconstructed Si(100)-2$\times$1 surface
and on top of the dimer rows.
An adatom initially sitting on top of a row was observed to move in between rows during STM manipulation, indicating that the latter site is 
more stable. 
Previous density functional theory (DFT) studies had reported binding sites 
in between the dimer rows (BDR) \cite{ju2008ab} and there it was assumed that diffusion occurs by adatom hops between such binding sites, 
the adatom thereby remaining in between dimer rows during diffusion.
More extensive DFT calculations by Chiaravalloti {\it et al.}, however, identified also two binding sites on top of the dimer rows,
an asymmetric site at the edge of the dimer row (TDR1) and a symmetric site in the center of a row (TDR2).
The question then arises how Au adatoms diffuse on the surface, in particular whether the TDR sites play some role there
or whether the diffusion occurs by direct hopping between the optimal BDR sites.   
This will, for example, affect where dimers form and how Au islands nucleate on the surface. 
 

In this article, results of theoretical calculations of the diffusion of a Au adatom on the reconstructed Si(100)-2x1 surface are presented. 
The study is based on nudged elastic band (NEB) calculations of minimum energy paths for elementary transitions 
with the energy and atomic forces estimated from DFT calculations with the PBEsol functional.
The results, combined with rate estimates based on harmonic transition state theory are used in kinetic Monte Carlo (KMC) 
simulations of Au adatom diffusion over a range of temperature. The optimal diffusion mechanism turns out to be 
non-intuitive and indirect involving Au adatom hopping on top of a dimer row and then migration along the row long distance between
visits to the low energy BDR sites. The effective activation energy turns out to be significantly smaller than that of 
the more direct diffusion path between BDR sites. 


\section{Methodology}

\begin{figure}[!b]
    \centering
    \includegraphics[width=0.6\textwidth]{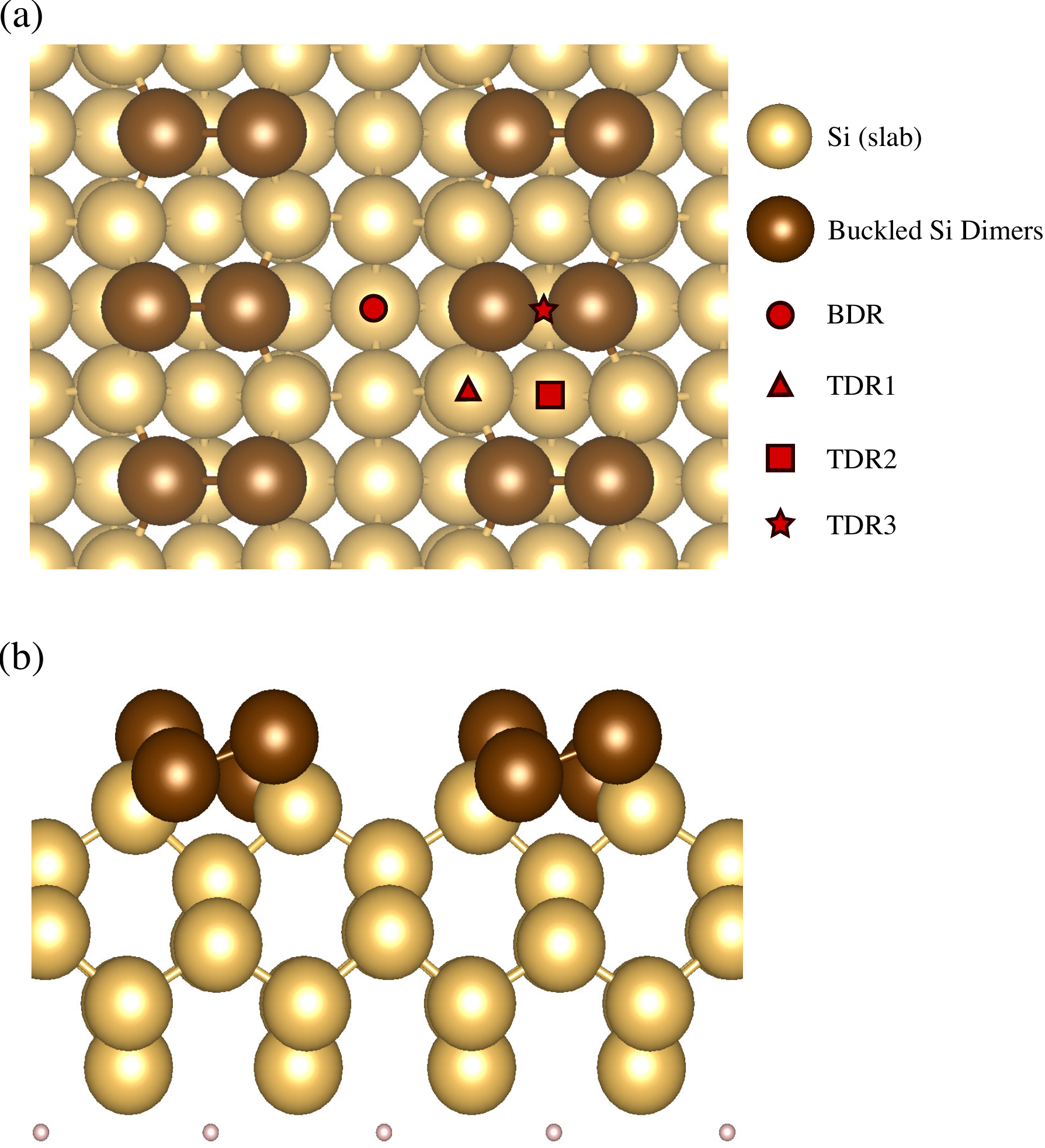}
    \caption{
    Top (a) and side (b) views of the clean Si(100)-2$\times$1 surface. The Si atoms in the buckled dimers are highlighted with a darker color and the capping hydrogen atoms are shown in lighter color. 
    The binding sites of the Au adatom, in between dimer rows (BDR, representing both BDR1 and BDR2, see Fig.~\ref{fig:BDRs}) 
    and on top of a dimer row (TDR1, TDR2 and TDR3), are marked
    with red symbols. }
    \label{fig:SiSlab_adsSites}
\end{figure}

The reconstructed Si(100)-2$\times$1 surface is modeled with a periodic 4$\times$4 surface supercell of a six layer slab. 
The three upper layers are allowed to relax while the rest of the atoms are kept frozen at the perfect crystal positions. 
The bottom silicon layer is passivated with hydrogen atoms. The system is illustrated in Fig.~\ref{fig:SiSlab_adsSites}.

To calculate the minimum energy paths between the local minima corresponding to the Au adatom binding sites, 
the climbing image NEB (CI-NEB) method is used~\cite{Henkelman_2000a,Henkelman_2000b,Asgeirsson2018b}. 
The image dependent pair potential (IDPP) method~\cite{Smidstrup_IDPP_2014} with six intermediate images is used to generate the initial paths. 
Iterative optimization of atomic coordinates is carried out until the magnitude of the components of the atomic forces 
perpendicular to the path have dropped below 0.01~eV/{\AA}.

The energy of the system and atomic forces are estimated using DFT within the PBEsol functional approximation \cite{Perdew_2008_pbesol}.
PBEsol is chosen here because it is known to give good results for the silicon crystal and its surface, including the surface energy. 
It is closer to the local density approximation (LDA) than the PBE approximation in that the exchange enhancement factor is smaller.
As a result, the energy of an isolated atom is higher with PBEsol and the binding energy to the surface therefore overestimated, but this
error will be similar for all binding sites on the surface and not affect the shape of the energy landscape for the adatom. 
A plane wave basis set is used with a cutoff energy of 350~eV to represent the valence electrons,
while the inner electrons are represented with the projector augmented-wave method \cite{Bloechl94}.
Calculations of the crystal give a lattice parameter of 5.43~\AA~ in excellent agreement with the experimental value.
A vacuum space of 15~\AA~is used to avoid interaction between periodic images of the slab. 
The Brillouin zone is sampled using a uniform mesh with 7$\times$7$\times$1 k-points.  
The calculations are carried out with the EON software~\cite{EON} with energy and atomic forces obtained from VASP~\cite{VASP}.

The values of the activation energy, $E_a$, for each elementary hop of the Au adatom from one binding site to another are 
obtained from the minimum energy paths as the maximum energy along the path minus the initial state energy. 
The rate constants for the various processes are then estimated using the Arrhenius expression,
$k = \nu \exp{(-E_a/k_BT)}$, where the pre-exponential factor is taken to have a typical value of $\nu$~= 10$^{12}$ s$^{-1}$.
The transition mechanisms and estimated rate constants are then used to prepare input for a KMC simulation of the diffusion over a 
larger area of the surface and a range of temperature values using the Zacros software \cite{Zacros_2011,Nielsen_2013}. 
The possibility for desorption or adsorption of gold atoms is not included as the goal here is to identify the possible diffusion paths of 
a single Au adatom.


\section{Results and Discussion}

\subsection{Binding sites}

It is well known that the Si(100) surface undergoes a reconstruction to form an extended 2$\times$1 surface unit cell where dimer rows are formed to reduce dangling bonds. We started our calculations from the unreconstructed Si(100) surface, with two dangling bonds for every surface Si atom. In Fig.~\ref{fig:SiSlab_adsSites} on-top and side views are shown of the clean Si(100)-2$\times$1 surface obtained after optimization
of the atomic coordinates using atomic forces evaluated from DFT/PBEsol. The distinctive buckling dimer rows 
are highlighted with a different color. The calculations give a dimer bond length of 2.34~\AA~and a buckling angle of 19.8$^\circ$. This agrees quite well with data obtained from low energy electron diffraction (LEED) experiments giving values of 2.24~\AA~and 19.2$^\circ$ 
 \cite{Over_1997}. 

At first, the three binding sites of the Au adatom identified in the DFT/PBE calculations of Chiaravalloti {\it et al.} \cite{chiaravalloti2009stm}
are calculated by placing the adatom in the vicinity of these locations and minimizing the energy with respect to the coordinates of all the 
movable atoms. These binding sites are labeled as BDR1, TDR1 and TDR2, as indicated in Fig.~\ref{fig:SiSlab_adsSites}. 
After the local minimum has been reached, the binding energy, $E_{b}$, is calculated as
\begin{equation}
    E_{b} = E_{surf}+E_{Au} - E_{Au/surf},
\end{equation}
where $E_{Au/surf}$ and $E_{surf}$ correspond to the energy of the silicon slab with and without the Au adatom, respectively, 
and $E_{Au}$ is the energy of an isolated Au atom. 
Table 1 shows the values obtained here with the PBEsol functional as well as the PBE values obtained previously by
Chiaravalloti {\it et al.} \cite{chiaravalloti2009stm}.
The two functionals give similar values, the PBEsol binding energy being larger by about 0.1 to 0.2 eV
as could be expected from the overestimate of the energy of the isolated Au atom.

At all the binding sites, the Au adatom is stabilized by bonding to two Si-dimer atoms. 
The Si-Au-Si angle formed is given in Table~\ref{tab:BindingE}. 
The BDR1 position represents the most stable adsorption site, consistent with the fact that it is the most frequently observed configuration in 
the low temperature STM measurements \cite{chiaravalloti2009stm}. 
The TDR1 and TDR2 sites appear to be nearly equally stable, with a binding energy difference of only 0.05~eV in the PBEsol calculations.

\begin{table}[!b]
\small
\begin{tabular}{lcc}
\hline\hline
     & DFT/PBEsol    & DFT/PBE$^a$       \\ \hline
BDR1  & 3.44 (140$^\circ$) & 3.24 (131$^\circ$)  \\
BDR2  & 2.91 ($170.5^\circ$) &   \\
TDR1 & 3.12 (100$^\circ$)  & 3.03 (98$^\circ$)  \\
TDR2 & 3.07 (121$^\circ$) & 2.94 (118$^\circ$) \\ 
TDR3 & 2.87   (58$^\circ$)&     \\ 
\hline\hline
\end{tabular}
\caption{
Binding energy (in eV) of a Au adatom at various adsorption sites on the Si(100)-2$\times$1 surface, calculated with
the PBEsol and PBE functional approximations. 
The Si-Au-Si angle is given in brackets (in deg.).\\
$^a$Data from Ref.~\cite{chiaravalloti2009stm}}
\label{tab:BindingE}
\end{table}

\begin{figure}[!b]
    \centering
    \includegraphics[width=0.6\textwidth]{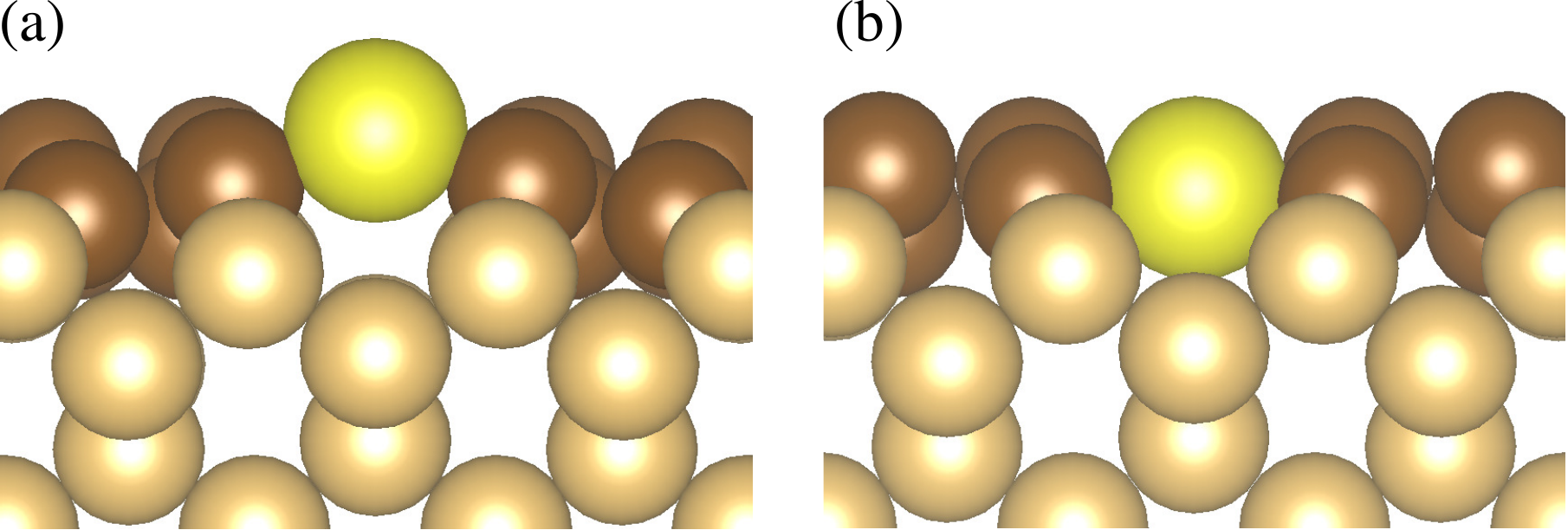}
    \caption{
    Side view of the two local minima found for the Au adatom placed in between silicon dimer rows (BDR sites). 
    The Au adatom is displayed in yellow and the Si-dimer atoms are colored in brown. 
    The Au adatom in the lower site, BDR2, shown in (b), is 1~\AA~closer to the surface than in the more stable upper site, BDR1, shown in (a).
    }
    \label{fig:BDRs}
\end{figure}

An additional local minimum in the BDR position has also been found (from now on referred to as BDR2) where the Au adatom is located 
$\approx$1~\AA~closer to the surface as compared to the previously known BDR1 configuration. 
Fig.~\ref{fig:BDRs} shows a side view comparison between these two configurations of the atoms. 
The calculated binding energy for the BDR2 configuration is 2.91~eV, 
about 0.5 eV smaller than the one obtained for the BDR1 site. 
There, the Au atom is again bonded to two Si-dimer atoms but forms a significantly larger Si-Au-Si angle of 170.5$^\circ$.


\subsection{Diffusion mechanism}

\begin{figure*}[t]
    \centering
    \includegraphics[width=0.99\textwidth]{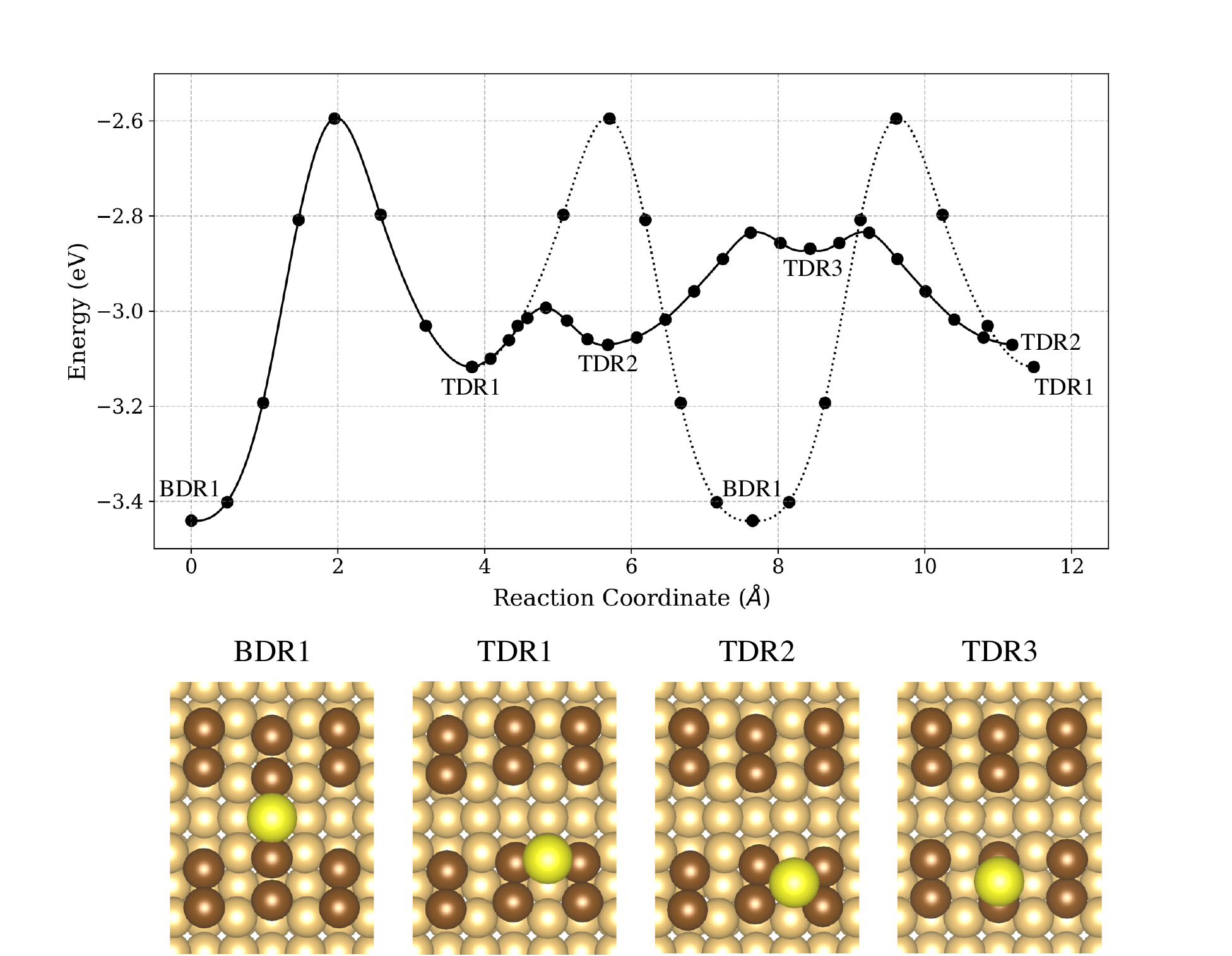}
    \caption{
    Calculated minimum energy paths between the main adsorption sites obtained using the CI-NEB method with 
     DFT/PBEsol atomic forces. 
    Each dot corresponds to an image of the system along the path. 
    Bottom panels correspond to on-top view of the BDR1, TDR1, TDR2 and TDR3 configurations.
    The TDR3 site was discovered as an intermediate minimum in the CI-NEB calculations of the minimum energy path between 
    adjacent TDR2 sites.
    }
    \label{fig:MEPs}
\end{figure*}

The mechanism of Au adatom diffusion on the surface is found by identifying the minimum energy paths connecting the binding sites.
The results are shown in Fig.~\ref{fig:MEPs}.
Starting with the adatom in one of the most stable binding sites, a BDR1 site, an initial path to an adjacent BDR1 site
is generated using the IDPP method.
The NEB optimization of the path results in a longer path that visits a TDR1 site as an intermediate minimum.
It turns out that there is no minimum energy path between
two BDR1 sites that does not include an intermediate minimum.
The calculated activation energy for the BDR1$\rightarrow$TDR1 hop is 0.84~eV, while the opposite process,
 to jump back to the same or an adjacent BDR1 site,
has an activation energy of 0.52~eV. 
Once the Au adatom is in a TDR1 site, it has, however, other options than to fall down to a BDR1 site.
A hop to a TDR2 site at the center of the dimer row has a lower activation energy of 0.12~eV.
From the TDR2 site, the adatom can either go back to the TDR1 site by overcoming a barrier of only 0.08~eV, 
or it can slide over a silicon dimer  to get to an adjacent TDR2 site. 
The CI-NEB calculation of the path between two TDR2 sites reveals an  
intermediate binding site where the adatom sits on top of a Si-dimer. 
This site is labeled as TDR3 in Figs.~\ref{fig:SiSlab_adsSites} and ~\ref{fig:MEPs}). 
The energy barrier for the  TDR2$\rightarrow$TDR3 hop is 0.24~eV.
The TDR2$\rightarrow$TDR2 process can also involve dissociation of the Si-dimer as the Au atom goes through it instead of going over it. 
However, this process has an activation energy of 0.57~eV, making it less likely.  

\begin{figure}[!h]
    \centering
    \includegraphics[width=0.45\textwidth]{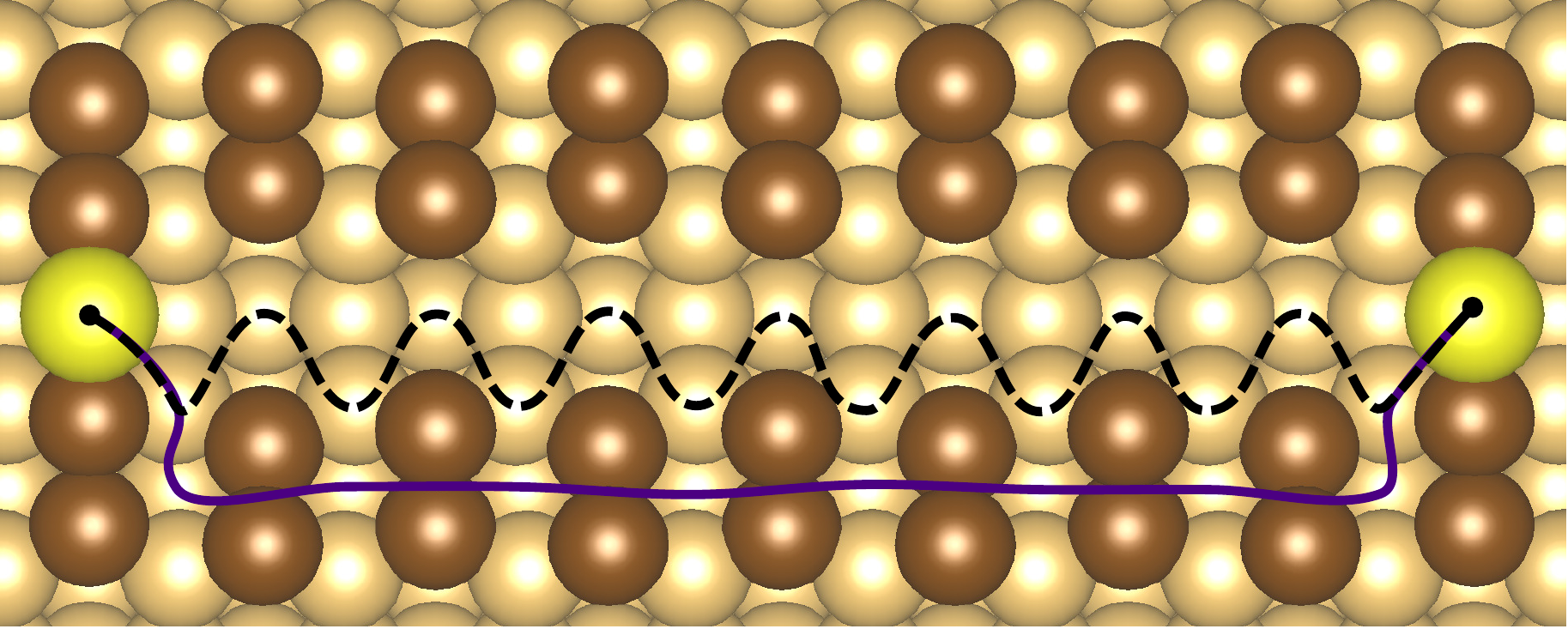}
    \caption{
    Illustration of the two competing diffusion paths for the Au adatom. 
    The dashed line represents the most direct diffusion path between optimal sites, BDR1$\rightarrow$TDR1$\rightarrow$BDR1. 
    The solid line indicates schematically the more efficient indirect path involving fast migration on top of a dimer row involving 
    TDR1, TDR2 and TDR3 sites before falling back down again into one of the optimal BDR1 binding sites.
    }
    \label{fig:diffusionScheme}
\end{figure}

Thus, the diffusion from one BDR1 site to another can occur via two possible paths (Illustrated in Fig.~\ref{fig:diffusionScheme}). 
In the most direct path, the Au adatom goes first to a TDR1 site and then to another BDR1 site.
Another, less direct path involves visiting also TDR2 sites from the TDR1 site and possibly extended travel along the silicon dimer row until 
the adatom eventually jumps to a BDR1 site. In order to study the competition between these different diffusion mechanisms,  
KMC simulations of the long time scale dynamics were carried out for a range of temperature. 

\section{Kinetic Monte Carlo simulations}

\begin{table*}[!t]
\small
\begin{tabular}{ccccc}
\hline\hline
T (K)  & $\tau_{BDR1}$   & $\tau_{TDRs}$ & L/d \\ 
\hline
200 & \num{3.5e+08}   &\ \num{2.7e-01} & 1910 \\
250 & \num{2.3e+04}  & \ \num{1.2e-03} & 240  \\
300 & \num{3.0e+01}  & \ \num{2.0e-05} & 72   \\
400 & \num{9.4e-03}  & \ \num{2.0e-07} & 19   \\
500 & \num{6.8e-05}  & \ \num{1.6e-08} & 11   \\
700 & \num{2.8e-07}  & \ \num{7.9e-10} & 5    \\
900 & \num{1.2e-08}  & \ \num{1.3e-10} & 3    \\ 
%
%
\hline\hline
\end{tabular}
\caption{
Results of kinetic Monte Carlo simulations at various values of the temperature:
$\tau_{BDR1}$ is the average time in seconds spent in a BDR1 site,  
$\tau_{TDRs}$ is the time spent on sites on top of dimer rows in between visits to 
BDR1 sites, and
L/d is the length traveled along the top of a dimer row in unit of BDR1 site separation, d= 3.84 \AA.
}
\label{tab:statsZacros}
\end{table*}

\begin{figure}[!h]
    \centering
    \includegraphics[width=0.45\textwidth]{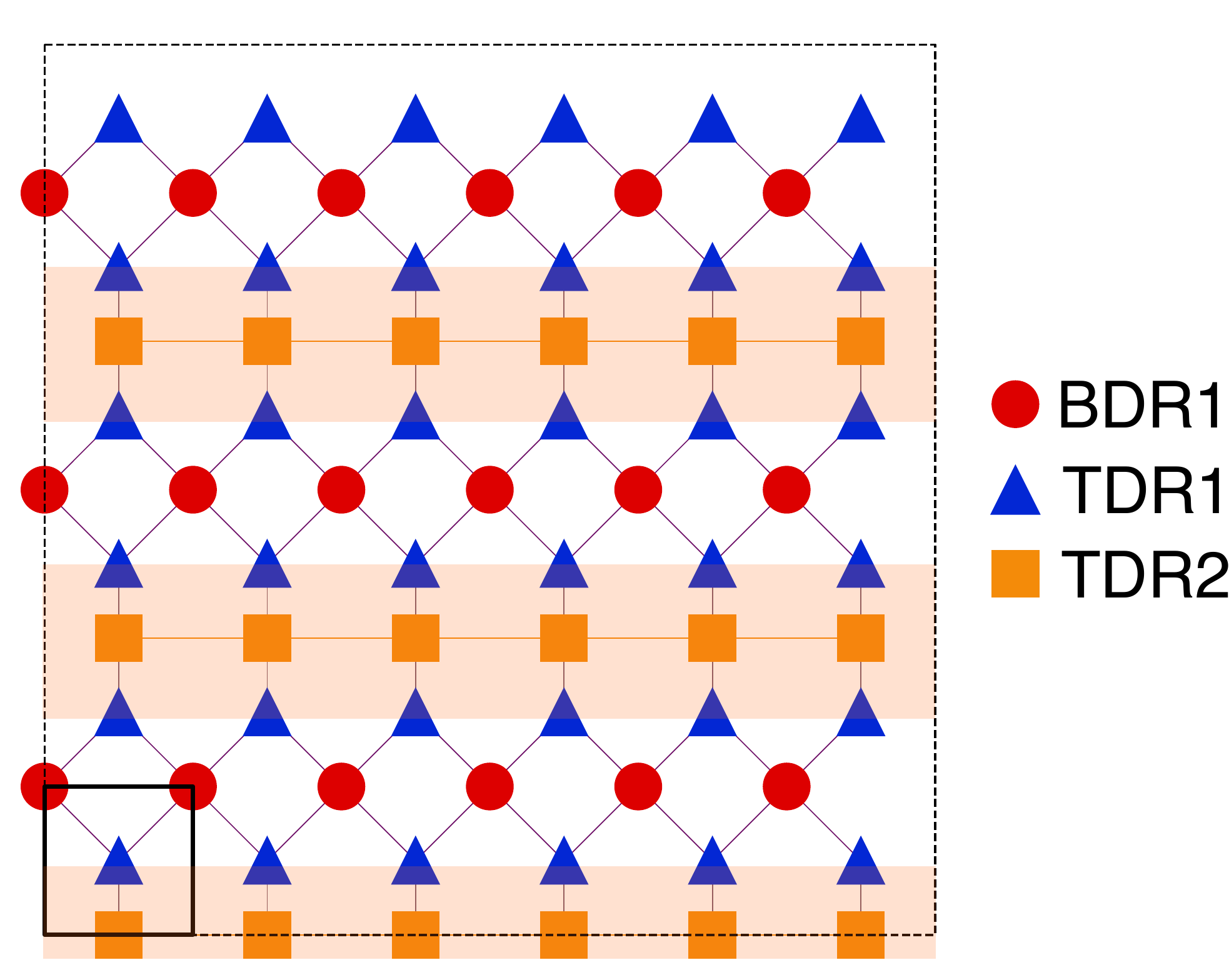}
    \caption{
Illustration of the lattice of sites included in the KMC simulations of the diffusion of the Au atom. 
For simplicity, the weakly binding TDR3 site is not included in the simulations.
    }
    \label{fig:kmclattice}
\end{figure}

The lattice used in the KMC simulations includes only the most relevant adsorption sites (\textit{i.e.} BDR1, TDR1 and TDR2),
as illustrated in Fig.~\ref{fig:kmclattice}.
The elementary transitions included correspond to the reversible adatom hops: 
(1) between BDR1 and TDR1; (2) between TDR1 and TDR2; and (3) between adjacent TDR2 sites. 
The activation energy for each hop is taken from the DFT/PBEsol calculations described above.
Adsorption or desorption of the Au atom is not included in the simulations. 
The initial configuration corresponds to a Au atom adsorbed in one of the BDR1 sites. 

A total of 10 simulations were carried out for seven values of the temperature: 200, 250, 300, 400, 500, 700 and 900~K.
For each simulation, a different random number seed was used to generate independent instances of time evolution of the system.
An average of $\num{5e+09}$ KMC events occurred in each simulation. 
Table~\ref{tab:statsZacros} presents the average time spent in a BDR1 site
as well as the average time spent on top of a dimer row and number of TDR2 sites visited in between BDR1 sites. 
The results show that the Au adatom tends to skid along a dimer row, especially at low temperature, rather 
than the BDR1$\rightarrow$TDR1$\rightarrow$BDR1 path.
This preference at low temperature stems from the fact that once the adatom has made it to a TDR1 site, the energy barrier for skidding 
along a dimer row is lower than the barrier for entering a BDR1 site.
At high temperature, this difference in barrier height is less important and the BDR1$\rightarrow$TDR1$\rightarrow$BDR1 path 
becomes more competitive. 

\begin{figure}[!t]
    \centering
    \includegraphics[width=0.6\textwidth]{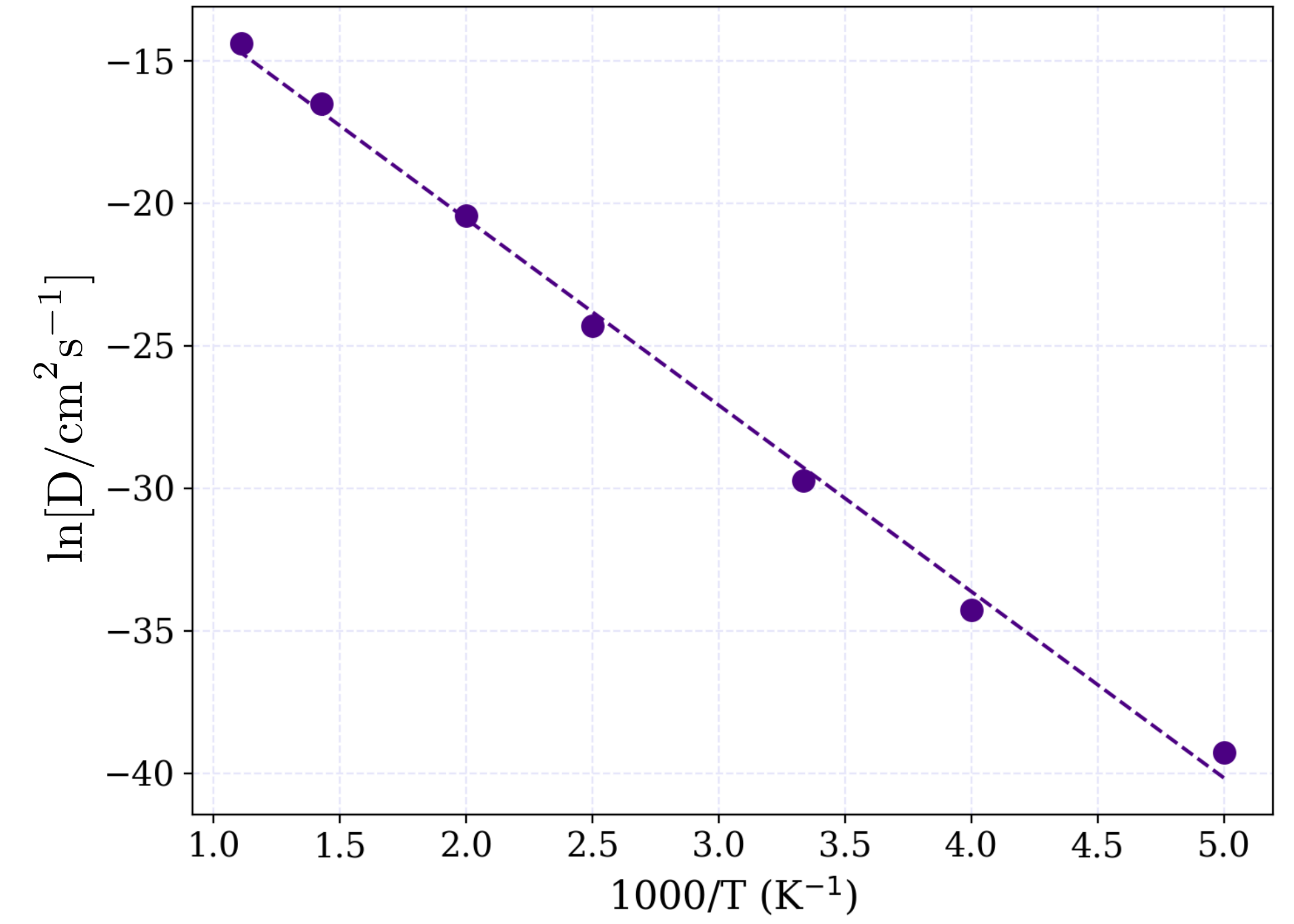}
    \caption{
Arrhenius graph of the Au adatom diffusion coefficient, $D$, on Si(100)-2$\times$1 as a function of temperature in the range 200~K to 900~K. 
The linear fit gives an effective activation energy of 0.56~eV.
There is, however, clear deviation from linear dependence because the indirect mechanism of skidding along a dimer row becomes 
more important as the temperature is lowered.   
    }
    \label{fig:ArrheniusFig}
\end{figure}

The KMC results give a value for the average distance traveled by the Au adatom along a dimer row in between visits to BDR1 sites.
This can be used to estimate the diffusion coefficient, $D$, assuming a one-dimensional random walk between BDR1 sites as 
\begin{equation}
D(T) = \frac{L^2}{2\tau}    
\end{equation}
where $L$ is the average length travelled and $\tau_{BDR1}$ is the average time spent at a BDR1 site before hopping on top of a dimer row. Fig.~\ref{fig:ArrheniusFig} shows the Arrhenius graph of this estimate of the diffusion coefficient. 
Since the skidding along a dimer row is more important relative to the more direct BDR1$\rightarrow$TDR1$\rightarrow$BDR1 path at
low temperature, the relationship is not linear.
A linear linear fit to the whole range from 200~K to 900~K gives an average slope corresponding 
to an effective activation energy for diffusion as 0.56~eV. 
The slope, however, is lower in the lower temperature range than in the high temperature range as the relative importance of the two 
diffusion mechanisms changes with temperature.
This value of the effective activation energy is significantly lower than the activation energy for the direct BDR1$\rightarrow$TDR1$\rightarrow$BDR1 path which is 0.84 eV.
Fig.~\ref{fig:diffusionScheme} shows a depiction of the competing diffusion paths.
The solid line indicates a simplified description of the indirect mechanism
where the adatom hops on top of a dimer row in rapid migration.
The dashed line represents the more direct BDR1$\rightarrow$TDR1$\rightarrow$BDR1 mechanism 
that becomes competitive only at high temperature.

The diffusion mechanism for the Au adatom on the Si(100)-2x1 surface identified here is similar in many respects to 
the diffusion mechanism of a Si adatom on this surface, in that an indirect diffusion mechanism involving repeated 
hops along of the top of a dimer row rows turn out to 
be more efficient than a direct hopping mechanism between optimal binding sites, especially at low temperature~\cite{Smith_Jonsson_PRL1996}.
This explained experimental STM observations of the formation of Si addimers on top of dimer rows 
while dimers were expected to form in between dimer rows since those sites have greater binding energy.


\section{Conclusion}

The mechanism and rate of diffusion of a Au adatom on the reconstructed Si(100)-2$\times$1 surface 
has been calculated using the CI-NEB method for identifying optimal diffusion paths with energy and atomic forces estimated from DFT/PBEsol.
While the most stable binding site is found to be in between dimer rows, in agreement with previous theoretical calculations and STM experimental 
measurements~\cite{ju2008ab,chiaravalloti2009stm}, the dominant diffusion mechanism is found to involve
promotion of the adatom into metastable sites on top of a dimer row and multiple hops along the row, 
before it settles down again into an optimal binding site between dimer rows. 
This indirect diffusion mechanism becomes more dominant the lower the temperature is. At room temperature the adatom is 
predicted to skid along a dimer row covering distances on the order of 300 \AA\
in between visits to optimal binding sites.

In addition to the optimal binding site in between dimer rows, an additional local minimum, BDR2, is found where the Au adatom is 
1~\AA~ closer to the surface but this site corresponds to higher energy by 0.5~eV. 
Also, a weak binding site on top of a Si dimer, the TDR3 site, has been identified as an intermediate
minimum in the CI-NEB calculation on the minimum energy path between adjacent TDR2 sites.
%
The Au adatom can also split a Si-dimer in order to pass through it during the transition between TDR2 sites, 
but the energy barrier is 0.57~eV so this process is less likely than a hop over the Si dimer.

Simulations of diffusion paths over a range of temperature using the KMC method reveal the relative importance of the indirect and direct diffusion 
paths and are used to estimate the diffusion coefficient. From the temperature dependence of the diffusion coefficient an effective activation
energy of 0.52 eV is obtained, significantly lower than the activation energy for the direct diffusion mechanism, 0.84 eV. 
%
The indirect diffusion mechanism can have significant consequences for the formation of dimers and larger Au islands on the surface. 
Since the adatoms travel long distances on top of dimer rows, the most probable site for the formation of a Au addimer is on 
top of a dimer row, rather than in between dimer rows as one would predict from the location of the optimal binding sites.
%


\begin{acknowledgements}

This project was funded by
European Union’s Horizon 2020 research and innovation programme under grant agreement No 814416, and
the Icelandic Research Fund. 
We thank Adam Foster for helpful discussions. 
The calculations were carried out at the Icelandic Research High Performance Computing (IRHPC) facility.

\end{acknowledgements}



\end{document}